\documentclass[prd,showpacs,nofootinbib,preprintnumbers,
11pt ]{revtex4}
\usepackage{amsmath} \usepackage{graphicx} \usepackage{amsfonts}
\usepackage{array} \usepackage{amsthm} \usepackage{bm} \usepackage{mathptmx}

\usepackage{latexsym}
\newcommand{\nn}{\nonumber}
\newcommand{\be}{\begin{equation}}
\newcommand{\ee}{\end{equation}}
\newcommand{\ba}{\begin{eqnarray}}
\newcommand{\ea}{\end{eqnarray}}
\newcommand{\bal}{\begin{align}}
\newcommand{\eal}{\end{align}}
\newcommand{\lb}{\label}

\newcommand{\fr}{\frac}

\newcommand{\bw}{\begin{widetext}}
\newcommand{\ew}{\end{widetext}}

\def\6{_{6}}
\def\5{_{5}}
\def\4{_{4}}

\def\A{{\cal A}}
\def\F{{\cal F}}

\newcommand{\G}{\hat{G}}

\newcommand{\hmu}{\hat{\mu}}
\newcommand{\hnu}{\hat{\nu}}
\newcommand{\hrho}{\hat{\rho}}
\newcommand{\hlambda}{\hat{\lambda}}
\newcommand{\hpsi}{\hat{\psi}}
\newcommand{\hsigma}{\hat{\sigma}}
\newcommand{\htau}{\hat{\tau}}

\begin{document}
\begin{flushright}LAPTH-001/13
\end{flushright}
\begin{flushright}DTP-MSU/12-13
\end{flushright}

\title{Oxidation of  $D=3$ cosets and Bonnor dualities in $D\leq 6$ }

\author
{G\'erard Cl\'ement}\email{gclement@lapp.in2p3.fr} \affiliation{
Laboratoire de  Physique Th\'eorique LAPTH (CNRS),
\\
B.P.110, F-74941 Annecy-le-Vieux cedex, France}
\author{Dmitri V. Gal'tsov} \email{galtsov@phys.msu.ru}
\affiliation{Department of Theoretical Physics, Moscow State
University, 119899, Moscow, Russia}

\date{\today}

\begin{abstract}
Bonnor's map in General Relativity  is duality between
(dimensionally reduced) vacuum gravity and static truncation of
electro-vacuum theory. It was used as a tool to generate an exact
solution of electro-vacuum from some vacuum solution. It can be
expected that similar dualities will be useful for solution
generation in higher-dimensional theories too. Here we study such
maps within a class of theories in dimensions $D\leq 6$ using
oxidation of $D=3$ cosets and consistent truncation of the
corresponding theories. Our class includes those theories whose
$D=3$ symmetries are subgroups of $G=O(5,4)$. It contains
six-dimensional minimal supergravity, five-dimensional minimal and
$U(1)^3$ supergravities and a number of four-dimensional theories
which attracted  attention recently in the search of exact
solutions. We give explicit reduction/truncation formulas relating
different theories in dimensions $D=4,\,5,\,6$ in terms of metrics
and matter fields  and discuss various alternative duality chains
between them.
\end{abstract}


\maketitle

\section{Introduction}

Toroidal dimensional reduction of bosonic sectors of supergravities
in $D\leq 11$ dimensions to $D=3$ leads to sigma models on coset
spaces \cite{desin,Cremmer:1997ct,Cremmer:1998px}. The $D$-
dimensional theories generically contain the metric and the scalar
(with no potentials), vector and higher-rank form fields. Assuming
the existence of $D-3$ commuting Killing vectors one reduces them to
three-dimensional theories with scalar and vector fields only, the
latter then being replaced by additional scalars using
three-dimensional Hodge dualisation. The final set of scalar fields
(in General Relativity usually called potentials) contains the same
number of degrees of freedom as the initial theory in $D$ dimensions
restricted to all fields depending only on $D-3$ coordinates. Due to
dualisation involved in this derivation the map between the initial
and the three-dimensional scalar variables is not point-like,
however. These scalars in many cases parameterize homogeneous spaces
$G/H$ where $G$ is some Lie group (semi-simple in most interesting
cases) and $H$ is the isotropy subgroup. A rather complete list of
such cosets was given by Breitenlohner, Maison and Gibbons
\cite{Breitenlohner:1987dg}, for more detailed discussion see
\cite{Maison:2000fj}).

The isometry groups $G$ of these cosets (often called hidden
symmetries) are  extremely useful in the search of exact solutions
in gravity/supergravity. Historically, the use of $D=3$ sigma models
for solution generation in four dimensions goes back to the papers
of Ehlers \cite{Ehlers:1957zz} (vacuum Einstein equations),
Neugebauer and Kramer \cite{nk} (electrovacuum), Maison
\cite{Maison:1979kx} and Cl\'ement \cite{Clement:1985gm,
Clement:1986bt,Clement:1986dn} (Kaluza-Klein theory). During the
past two decades similar methods were developed for various
supergravity/string effective actions in four
\cite{emda,Bakas:1996dz, Gal'tsov:1996gv,Gal'tsov:1996cm,
Gal'tsov:1997kp,Chen:1999rv,Clement:2001gia, bremda,Clement:2002mb}
and five dimensions
\cite{Bouchareb:2007ax,Gal'tsov:2008nz,5to3,Clement:2008qx,phantom2,AzregAinou:2011gt}.
Special attention was paid to the description of BPS states as
nilpotent orbits of coset isometries
\cite{Clement:1985gm,Clement:1986bt,Clement:1996nh,
Bossard:2009we,Bellucci:2009qv,Fre:2011uy}. A particular interest in
$D=5$ theories was motivated by the discovery of black rings and
related objects. The conjectured existence of similar solutions in
$D=6$ stimulates further investigation of $D=6$ theories along the
same lines.

Apart from using symmetries of a {\em given theory}, generating
techniques can be extended using maps between {\em different
theories} described by the same cosets, or their consistent
truncations. A simple example is provided by Bonnor's map
\cite{Bonnor} in General Relativity, relating the static subspace of
$D=4$ Einstein-Maxwell ($EM$)  theory and vacuum Einstein gravity,
both having the same hidden symmetry $SL(2,R)$. By mapping the
target space potentials related to these two theories, Bonnor was
able to derive an exact static magnetic dipole solution of $EM$
theory. Similar methods were applied to Kaluza-Klein theory
\cite{Clement:1986dn} and to Einstein-Maxwell-dilaton theory with
arbitrary dilaton coupling \cite{Galtsov:1995mb}.

A convenient framework to reveal Bonnor-type maps in more general
cases and higher dimensions is provided by the theory of oxidation
--- determination of  higher-dimensional theories whose toroidal compactification
to three dimensions leads to a given coset. The basic features of
oxidation  were described by Cremmer, Julia, Lu and Pope
\cite{Cremmer:1999du}, an extensive discussion from the
group-theoretical viewpoint was later given by Keurentjes
\cite{Keurentjes:2002xc,Keurentjes:2002rc}. Generically a coset can
be oxidized to various higher dimensions up to an ultimate oxidation
endpoint which can be identified using some  rules. One such rule
follows from the fact that $D$-dimensional vacuum Einstein gravity
leads  to the coset $SL(D-2, R)/O(D-2)$
\cite{Maison:1979kx,Cremmer:1998px}. Conversely, a maximal oxidation
of the $SL(n, R)/O(n)$ coset is ($D=n+2$)- dimensional General
Relativity. As was observed in \cite{Cremmer:1999du}, a similar rule
holds for  the maximally non-compact forms of $D_n$ ($G=O(n,n)$),
and $B_n$ ($G=O(n+1,n)$), with an exception for $B_3$, whose
standard oxidation endpoint is $D=5$, but which can be further
oxidized to $D=6$ leading to minimal $D=6$ supergravity with a
self-dual three-form. It was also argued that the $C_n$ sequence
($G=Sp(2n, R)$) as well as some of the non-maximally non-compact
forms of $G=O(p,q)$ and $G=SU(p,q)$ cannot be oxidized beyond four
dimensions. It can be expected that lower-dimensional theories in
the oxidation chain of a given coset are related by toroidal
reduction of the theory corresponding to the oxidation endpoint
highest-dimensional member. In many cases this reduction is
accompanied by some non-trivial rearrangements of the matter fields.

The list of dualities is substantially enhanced if we include
oxidations of {\em invariant subspaces} of a given  coset, and this
is our basic idea. Actually, in the original Bonnor's case it was
duality between $D=4$ vacuum gravity  and the {\em static
truncation} of $D=4$ Einstein-Maxwell theory. Theories corresponding
to different subspaces are related to the parent theory by
consistent truncations. Therefore at each oxidation level we obtain
a chain of truncations which in turn can be reduced to an extended
set of lower dimensional theories. Mappings between different
members of the whole set of theories constitute the generalized
Bonnor-type dualities, which we investigate in this paper.  In
contrast with the original Bonnor map, they describe relations
between theories in different space-time dimensions.

\section{The setup}
Our goal  is to systematically study  the mappings between different
theories obtained  by oxidation of $D=3$ cosets which can be
embedded into $O(5,4)/(O(5)\times O(4))$ whose oxidation endpoint is
$D=6$. It is worth noting that this is not the largest coset
oxidizing  up to $D=6$: the coset of the exceptional group $F_4$,
also having the oxidation endpoint $D=6$, is larger. Rather, our
choice is motivated by the simplicity (and at the same time
multiplicity) of the resulting duality chains. More specifically,
our list of theories includes the following:
\begin{itemize}
\item
Heterotic string effective actions \cite{Hassan:1991mq} restricted
to $p$ vector fields, alternatively called
Einstein-Maxwell-Dilaton-Axion ($EM_p DA$) theories
 \be {\cal L}  =
\sqrt{g }\left(R - \frac12(\partial\psi)^2 - \frac14e^{\beta\psi}
{F}^a_{\mu\nu} {F}^{a\mu\nu} - \frac1{12}e^{2\beta\psi}
{G}_{\mu\nu\rho}\hat{G}^{\mu\nu\rho}\right)\,.
 \ee
with \be
 {F}^a_{\mu\nu} = 2 {A}^a_{[\nu,\mu]}\,, \quad  {G}_{\mu\nu\rho} =
3( {B}_{[\mu\nu\rho]} + \frac12 {F}^a_{[\mu\nu} {A}^a_{\rho]})\,,
\ee where sum over $a$ from one to $p$ is understood in all terms
quadratic in the vector fields, and the constant $\beta$ depends on
dimension. Our set will include the cases $p=0$ ( no vector fields,
$EDA$ ), $p=1,\; (EMDA),\; p=2 \;(EM_2 DA)$ and $p=3\; (EM_3 DA)$.
\item Minimal supergravities ($MSG$), whose bosonic lagrangians  are
 \be
{\cal L}\6 = \sqrt{g\6}\left( {R} - \frac1{12} {G}_{\mu\nu\lambda}
 {G}^{\mu\nu\lambda}\right)\,,
 \ee
in six dimensions,
 \be
{\cal L}\5 = \sqrt{g_5} \bigg(R - \frac14F^{\mu\nu}F_{\mu\nu}\bigg)
- \frac1{12\sqrt3}\epsilon^{\mu\nu\rho\sigma\lambda}F_{\mu\nu}
F_{\rho\sigma}A_{\lambda}\,,
 \ee
 in five, and the pure Einstein term in four dimensions.
\item $U(1)^3$-supergravity in $D=5$:
\be {\cal L}  =  \sqrt{g_5}\left(R - \frac12 {\cal G}_{IJ}(\partial
X^I)(\partial X^J) - \frac14 {\cal G}_{IJ} F^I_{\mu\nu }F^J_{\mu\nu
}\right)-\frac1{24} \epsilon^{\mu\nu\lambda\rho\tau} {\cal
\delta}_{IJK} F^I_{\mu\nu} F^{J\lambda\rho}A^K_\tau,  \ee where the
three scalars $X_I, \, I=1,2,3,$ are constrained by $X_1X_2X_3=1$,
$F^I=dA^I$, and  ${\cal G}_{IJ}={\rm
diag}\left((X^1)^{-2},(X^2)^{-2},(X^3)^{-2}\right),\;{\cal
\delta}_{IJK} =1$ for the permutation of the indices $1,2,3$ and
zero otherwise.
\item Einstein-Maxwell ($EM$) theory in $D=4$.
\item Vacuum Einstein gravity ($VG$),
\end{itemize}
Their hidden symmetry groups $G$  are listed in the Table
\be\lb{list}
\begin{tabular}{|c|c|c|c|}\hline
  &D=6 &D=5 &D=4\\ \hline
$VG$& $SL(4,R)$ &$SL(3,R)$ &$SL(2,R)$\\\hline $EDA$ & $O(4,4)$
&$O(3,3)$ &$O(2,2)$\\\hline $EMDA $&$O(5,4)$ &$O(4,3)$ &$O(3,2) \sim
Sp(4,R)$\\\hline $EM_2DA$&
---& $ O(5,3)$&$ O(4,2)\sim SU(2,2)$\\\hline $EM_3DA$ &---  &---
&$O(5,2)$ \\\hline $MSG$ & $O(4,3)$ &$G_{2,(2)}$ &$SL(2,R)$\\\hline
$U(1)^3\; SG$ &
--- & $O(4,4)$ &--- \\\hline $EM$& ---& --- &$SU(2,1)$\\\hline
\end{tabular}
\ee Empty spaces for $EM$ theories correspond to the absence of
non-trivial hidden symmetries. Empty spaces for the $EM_pDA$ series
for $D=5,\,6$ correspond to $G$ beyond the ultimate group $O(5,4)$,
namely, $EM_2DA$ in $D= 6$ has $G=O(6,4)$, while  $EM_3DA$  has
$G=O(7,4)$ in $D= 6$ and $G=O(6,3)$ in $D= 5$.

These theories are related by duality mappings of two sorts. The
first refers to different oxidation points of the same coset. Such
theories are related by toroidal dimensional reduction from the
oxidation endpoint. The second  refers to theories whose hidden
symmetries are subgroups of the largest group. These are connected
by consistent truncations of the largest theory. As a rule for
consistent truncations we use a slightly modified version of that
given in \cite{Breitenlohner:1987dg}:

{\em Proposition.} Let ${\cal G}=G/H$ where $G$ is a simple group
and $H$ is its maximal compact subgroup\footnote{For definiteness we
consider compactification on a purely space-like torus. Reducing on
the time direction leads to non-compact forms of $H$ without
changing the desired maps. We will also write the space-time measure
as $\sqrt{g} d^Dx$, including a minus sign into the definition of
$g$.}. Then if $G_1\subset G$ is a subgroup, and $H_1\subset H$ is
the maximal compact subgroup of $G_1$, one has

i) ${\cal G}_1=G_1/H_1$ is a totally geodesic subspace of ${\cal
G}$;

ii) the $D$-dimensional oxidation of ${\cal G}$ can be consistently
truncated to ${\cal G}_1$.

Combining toroidal dimensional reduction and consistent truncations,
one is able to derive all the listed theories from the
six-dimensional (one vector) heterotic effective action  ($EMDA$)
with hidden symmetry $G=O(5,4)$. This theory admits the sequence of
consistent  truncations first to ($EDA$) with $G=O(4,4)$, then to
minimal six-dimensional supergravity  ($MSG$) with   $G=O(4,3)$ and
finally to vacuum gravity with $G=SL(4,R)$. We then perform
dimensional reduction of this set of theories to five dimensions.
The ``parent'' theory $G=O(5,4)$ compactified on $S^1$ goes beyond
our list (containing three vectors, a two-form and three scalars),
but it can be consistently truncated to $U(1)^3-$ five-dimensional
supergravity with $G=O(4,4)$ and to $EM_2DA$ with $G=O(5,3)$, giving
rise to two further chains of truncations. One chain goes through
$U(1)^3\; SG$ which further truncates to $EMDA$ with $G=O(4,3)$, and
then to either to $EDA$ with $G=SL(4,R)\sim O(3,3)$, or to minimal
supergravity with $G=G_{(2,2)}$, which can be further truncated only
to vacuum gravity with $G=SL(3,R)$. The other chain goes through
$EM_2DA$, meeting with the first one at the $EMDA$ level.
Compactification of these theories on $S^1$ and their consistent
truncations in four dimensions complete our list (\ref{list}).

Many of the above relations between the mentioned theories were
discussed earlier, most notably in
\cite{Cremmer:1998em,Cremmer:1999du}. Our goal here is to give
explicit realizations in terms of original field variables, aiming
to open new possibilities for solution generation technique.

\setcounter{equation}{0}
\section{Truncations in six dimensions}
We start with the largest $D=3$ coset $O(5,4)/(O(5)\times O(4))$
admitting $D=6$ as oxidation endpoint. It is a 20-dimensional
homogeneous space. The corresponding six-dimensional theory is the
heterotic string effective action with one vector field. The  set of
fields contain the metric $\hat{g}_{\hmu\hnu}$, the three-form field
$ \hat{G}_{\hmu\hnu\hrho}$ (axion), the Maxwell two-form
$\hat{F}_{\hmu\hnu}$ and the dilaton $\hpsi$, so we call this theory
EMDA  (the full heterotic string $U(1)^{16}$ effective action
toroidally reduced to six dimensions contains 24 vectors). The
Lagrangian reads \be\lb{EMDA6} {\cal L}^6_{EMDA} =
\sqrt{\hat{g}}\left(\hat{R} - \frac12(\partial\hpsi)^2 -
\frac14e^{\beta\hpsi}\hat{F}_{\hmu\hnu}\hat{F}^{\hmu\hnu} -
\frac1{12}e^{2\beta\hpsi}\hat{G}_{\hmu\hnu\hrho}\hat{G}^{\hmu\hnu\hrho}\right)\,.
\ee ($\beta^2 = 1/2$) with \be \hat{F}_{\hmu\hnu} =
2\hat{A}_{[\hnu,\hmu]}\,, \quad \hat{G}_{\hmu\hnu\hrho} =
3(\hat{B}_{[\hmu\hnu,\hrho]} +
\frac12\hat{F}_{[\hmu\hnu}\hat{A}_{\hrho]})\,. \ee

 This theory admits consistent truncations according to the chain of
subgroups
 \be
O(5,4) \longrightarrow O(4,4)  \longrightarrow O(4,3)
\longrightarrow O(3,3) \sim SL(4,R)\,,
 \ee
labeling the sequence of invariant subspaces of the initial coset:
\begin{itemize} \item $O(4,4)/(O(4)\times O(4))$ (dimension 16),
\item $O(4,3)/(O(4)\times O(3))$ (dimension 12),
\item $SL(4,R)/(O(3)\times O(3))$ (dimension 9).
\end{itemize}
All of them admit $D=6$ as the oxidation endpoint (the case
$O(4,3)/(O(4)\times O(3))$ being exceptional \cite{Cremmer:1999du}).
By dimensionality, it is clear that the first coset correspond to
setting zero the Maxwell field $\hat{F}_{\hmu\hnu}=0$ which under
reduction to $D=3$ generates four scalars. It is a consistent
truncation of the Lagrangian (\ref{EMDA6}), as can be easily seen
from the equations of motion. The truncated theory is $EDA$:
\be\lb{EDA6} {\cal L}^{\,6}_{EDA} = \sqrt{\hat{g}}\left(\hat{R} -
\frac12(\partial\hpsi)^2 -
\frac1{12}e^{2\beta\hpsi}\hat{G}_{\hmu\hnu\hrho}\hat{G}^{\hmu\hnu\hrho}\right)\,,
\ee
 where $\hat{G}=d\hat{B}$.

 Its further  truncation is clear from
the field equations \ba \nabla^2\hpsi &=&
\frac{\beta}{6}e^{2\beta\hpsi}\hat{G}^{\hmu\hnu\hrho}\hat{G}_{\hmu\hnu\hrho}
\,,\lb{sc}
\\
D_{\hrho}(e^{2\beta\hpsi}\hat{G}^{\hmu\hnu\hrho}) &=& 0\,, \quad
D_{\hrho} \tilde{\hat{G}}^{\hmu\hnu\hrho} = 0\,. \ea These are
consistent with setting the dilaton to zero and imposing a
self-duality constraint on the three-form field, in which case the
right hand side of Eq. (\ref{sc}) vanishes:
 \be\lb{trunc44to43}
\hpsi = 0\,, \quad  \hat{G}_{\hmu\hnu\hrho}=
\tilde{\G}_{\hmu\hnu\hrho}\,.
 \ee
This gives the bosonic lagrangian of six-dimensional minimal
supergravity : \be\lb{MSG6} {\cal L}^{\,6}_{MSG} =
\sqrt{\hat{g}}\left(\hat{R}  - \frac1{12}
\hat{G}_{\hmu\hnu\hrho}\hat{G}^{\hmu\hnu\hrho}\right)\,,
 \ee
 with the three-form field
 \be \hat{G}_{\hmu\hnu\hlambda} \equiv 3\hat{B}_{[\hmu\hnu,\hlambda]}
\ee constrained by the self-duality condition \be
\hat{G}_{\hmu\hnu\hlambda} = \frac16\sqrt{\hat{g}}
\epsilon_{\hmu\hnu\hlambda\hrho\hsigma\htau}\hat{G}^{\hrho\hsigma\htau}\,.
 \ee
Toroidal reduction of this theory to three dimensions leads to the
coset $O(4,3)/(O(4)\times O(3))$, for which it is an anomalous
oxidation point, as was pointed out in \cite{Cremmer:1999du}.

Finally this theory can be truncated to the six-dimensional vacuum
Einstein theory, which is an ultimate oxidation point of the coset
for the group $SL(4,R) \sim O(3,3)$.   To summarize, the
six-dimensional truncation sequence is
 \be
EMDA \longrightarrow EMD  \longrightarrow MSG  \longrightarrow VG\,.
 \ee

\setcounter{equation}{0}
\section{Reduction to $D=5$}
The above six-dimensional  theories being compactified on $S^1$ give
rise to five-dimensional theories with the same hidden symmetries.
The parent theory $O(5,4)/(O(5)\times O(4))$ leads to a
five-dimensional theory containing three vectors, three scalars and
a two-form field. This is beyond our list, but can be further
truncated to other $D=5$ theories according to the scheme
 \be\lb{chain5}
\begin{array}{cc}    O(5,4)
 \\ \swarrow \qquad \searrow \\ O(4,4)\qquad O(5,3) \\
  \searrow \qquad \swarrow \\O(4,3)\\ \swarrow \qquad \searrow \\
G_{2(2)}\qquad O(3,3)\\\searrow \qquad \swarrow \\SL(3,R)
  \end{array}\quad\quad\qquad \begin{array}{cc}    \{3A_\mu,\;3S,\;
  B_{\mu\nu}\}
 \\ \swarrow \qquad \searrow \\ U(1)^3 \;SG\qquad EM_2DA \\
  \searrow \qquad \swarrow \\EMDA\\ \swarrow \qquad \searrow \\
MSG\qquad EDA\\\searrow \qquad \swarrow \\VG
  \end{array}
 \ee
Apart from the cosets listed in the previous section this includes
\begin{itemize} \item $O(5,3)/(O(5)\times O(3))$ (
dimension 15),
\item $G_{2(2)}/(SU(2)\times SU(2))$ (dimension 8),
\item $SL(3,R)/(O(3)\times O(3))$ (dimension 5).
\end{itemize}
Actually, the right subchain via the $EM_2DA$ sequence simply
corresponds to successive dropping of vector fields and we do not
give the details here. The left subchain via  $U(1)^3 \;SG$ is
rather non-trivial and can be alternatively regarded as
compactification of the truncated $D=6$ theories with
$G=O(4,4)\;(EDA)$ and $G=O(4,3)\;(MSG)$.

\subsection{Reducing $D=6$ EDA to $D=5\; U(1)^3$ supergravity}
Starting with the EDA model in six dimensions (\ref{EDA6})  and
reducing relative to a Killing vector $\partial_z$,
 \ba\lb{metric65}
ds\6^2 &=& e^{2\alpha\phi}g_{\mu\nu}dx^{\mu}dx^{\nu} +
e^{-6\alpha\phi} (dz+\A_{\mu}dx^{\mu})^2\,, \\ \lb{form65} \hat{B}
&=& \frac12B_{\mu\nu}dx^{\mu}\wedge dx^{\nu} +  A_{\nu}
dx^{\nu}\wedge dz
 \ea
where $\alpha^2=1/24$ and $\mu,\,\nu$ are five-dimensional indices,
we obtain
 \ba
{\cal L}\5 &=& \sqrt{g\5}\Big(R - \frac12(\partial\phi)^2 -
\frac12(\partial\psi)^2
-\frac14e^{-8\alpha \phi}\F_{\mu\nu}\F^{\mu\nu}-\nn\\
&-&\frac14e^{2\beta\psi+4\alpha \phi}F_{\mu\nu}F^{\mu\nu} -
\frac1{12}e^{2\beta\psi-4\alpha
\phi}G_{\mu\nu\rho}G^{\mu\nu\rho}\Big)\,,
 \ea
where $F=dA,\; G=dB-F\wedge \A.$ Variation over $B$ gives
 \be\lb{B}
D_\mu\left(e^{2\beta\psi-4\alpha \phi} G^{\mu\nu\rho}\right)=0,
 \ee
which is solved by
 \be\lb{G}
G^{\mu\nu\rho}= e^{-2\beta\psi+4\alpha
\phi}\frac1{2\sqrt{g_5}}\epsilon^{\mu\nu\rho\sigma\tau}K_{\sigma\tau}=e^{-2\beta\psi+4\alpha
\phi}{\tilde{K}}^{\mu\nu\rho},
 \ee with $K=dC$. Other equations
with account for (\ref{B},\ref{G}) read  \be D_\mu\left(
e^{2\beta\psi+4\alpha\phi}F^{\mu\nu}\right)=-\frac12
\tilde{K}^{\nu\mu\rho}\F_{\mu\rho}, \ee \be D_\mu\left( e^{-8\alpha
\phi}\F^{\mu\nu}\right)=-\frac12
 \tilde{K}^{\nu\mu\rho}F_{\mu\rho}. \ee The Bianchi identity
for $G$ can be rewritten similarly: \be D_\mu\left(
e^{-2\beta\psi-4\alpha \phi}K^{\mu\nu}\right)=-\frac12
 \tilde{F}^{\nu\mu\rho}\F_{\mu\rho}. \ee
The scalar equations are \ba\nabla^2\phi = -2\alpha
e^{-8\alpha\phi}\F_{\mu\nu}\F^{\mu\nu} + \alpha
e^{2\beta\psi+4\alpha\phi} F_{\mu\nu}F^{\mu\nu}-\frac{\alpha}3
e^{2\beta\psi-4\alpha\phi} G_{\mu\nu\rho}G^{\mu\nu\rho}\,,  \\
\nabla^2\psi =
\frac{\beta}2e^{2\beta\psi+4\alpha\phi}F_{\mu\nu}F^{\mu\nu} +
\frac{\beta}6e^{2\beta\phi-4\alpha\phi}G_{\mu\nu\rho}G^{\mu\nu\rho}\,,\ea
and the Einstein equations read
 \be R_{\mu\nu} =
\frac12\left\{\partial_{\mu}\phi
\partial_{\nu}\phi + \partial_{\mu}\psi
\partial_{\nu}\psi+ e^{-8\alpha\phi}\F_{\mu\lambda}{\F_{\nu}}^{\lambda}+
 e^{2\beta\psi+4\alpha\phi}F_{\mu\lambda}{F_{\nu}}^{\lambda}
+\frac12e^{2\beta\psi-4\alpha\phi}G_{\mu\lambda\rho}{G_{\nu}}^{\lambda\rho}
 \right\}.
  \ee
Define three scalars $X_I, \, I=1,2,3,$ constrained by $X_1X_2X_3=1$
as
 \be
 \ln X_1=\beta\psi-2\alpha\phi,\quad
 \ln X_2=-\beta\psi-2\alpha\phi,\quad
 \ln X_3=4\alpha\phi,\quad
 \ee
and denote $F^I_{\mu\nu}=(K_{\mu\nu},\,F_{\mu\nu},\, \F_{\mu\nu} )$
with potentials $F^I=dA^I$. Then the above system of equations can
be derived from the action of $U(1)^3\, D=5$ supergravity: \be {\cal
L}^5_{U3SG}  =  \sqrt{g\5}\left(R - \frac12 {\cal G}_{IJ}(\partial
X^I)(\partial X^J) - \frac14 {\cal G}_{IJ} F^I_{\mu\nu }F^J_{\mu\nu
}\right)-\frac1{24} \epsilon^{\mu\nu\lambda\rho\tau} {\cal
\delta}_{IJK} F^I_{\mu\nu} F^{J\lambda\rho}A^K_\tau,  \ee where
${\cal G}_{IJ}={\rm
diag}\left((X^1)^{-2},(X^2)^{-2},(X^3)^{-2}\right),\;{\cal
\delta}_{IJK} =1$ for the permutation of the indices $1,2,3$ and
zero otherwise.

\subsection{Reducing self-dual $D=6$ supergravity}
Now we start with the six-dimensional theory (\ref{MSG6}). The
six-dimensional Einstein equations read:
 \be \hat{R}^{\hmu}_{\hnu}
- \frac12\delta^{\hmu}_{\hnu}\hat{R} = \frac14
\hat{G}^{\hmu\hrho\hsigma}\hat{G}_{\hnu\hrho\hsigma}\,.
 \ee
Note that $\hat{G}^{\htau\hrho\hsigma}\hat{G}_{\htau\hrho\hsigma}=0$
for a self-dual 3-form\footnote{The lagrangian (\ref{MSG6}) itself
does not imply the self-duality condition, which should be imposed
by hand after the variation is performed. This is enough for our
purposes since actually we deal with the classical equations of
motion. For a consistent Lorentz-covariant lagrangian formulation of
theories involving chiral form fields see \cite{sorokin}. We thank
Dmitri Sorokin for discussion about this point.}.

To reduce to five dimensions we use the same ansatz (\ref{metric65})
for the metric and slightly modify the decomposition of the two-form
potential \be \hat{B}  =  \frac12B_{\mu\nu}dx^{\mu}\wedge dx^{\nu} +
\frac1{\sqrt2}A_{\nu} dz\wedge dx^{\nu}
 \ee
 leading to \be \hat{G}_{\mu\nu\lambda} = 3B_{[\mu\nu,\lambda]}\,,
\quad \hat{G}_{\mu\nu6} = \sqrt2A_{[\mu,\nu]} \equiv -
\frac1{\sqrt2}F_{\mu\nu}\,. \ee Self-duality of $\hat{G}$ gives \be
\hat{G}^{\mu\nu\lambda} =
\frac1{2\sqrt{2}\sqrt{g\5}}e^{-2\alpha\phi}
\epsilon^{\mu\nu\lambda\rho\sigma}F_{\rho\sigma} \equiv
e^{-2\alpha\phi} \frac1{\sqrt{2}}\tilde{F}^{\mu\nu\lambda}\,, \ee so
the six-dimensional equation
${\hat{G}^{\mu\nu\lambda}}_{;\lambda}=0$ is equivalent to the
five-dimensional identity \be D_{\lambda}\tilde{F}^{\mu\nu\lambda}
\equiv 0\,. \ee Also, \be - F_{\mu\nu} = -
\frac12\sqrt{g\5}e^{-4\alpha\phi}
\epsilon_{\mu\nu\rho\sigma\tau}(\sqrt{2}B^{[\rho\sigma,\tau]} +
\A^{\rho} F^{\sigma\tau})\,. \ee leading to \be
D_{\nu}(e^{4\alpha\phi}F^{\mu\nu}) =
\frac12\F_{\nu\rho}\tilde{F}^{\mu\nu\rho} =
\frac12e^{8\alpha\phi}F_{\nu\rho}H^{\mu\nu\rho}\,,\ee with \be
\F_{\mu\nu} \equiv 2\A_{[\nu,\mu]}\,, \quad H^{\mu\nu\rho} \equiv
-e^{-8\alpha\phi}\tilde{\F}^{\mu\nu\rho}\,. \ee These last
definitions imply \be D_{\lambda}(e^{8\alpha\phi}H^{\mu\nu\lambda})
= 0\,.  \ee Now one uses the mixed components of the Einstein
equations: \ba \hat{R}^{\mu}_6 &\equiv&
\frac12e^{-2\alpha\phi}D_{\nu}(e^{-8\alpha\phi} \F^{\mu\nu}) \nn
\\ &\equiv& \frac1{12\sqrt{g\5}}e^{-2\alpha\phi}
\partial_{\nu}(\epsilon^{\mu\nu\rho\sigma\tau}H_{\rho\sigma\tau}) \nn
\\ &=& -\frac18e^{-2\alpha\phi}F_{\nu\rho}\tilde{F}^{\mu\nu\rho}\,,\ea
implying \be H_{\mu\nu\rho} = 3(C_{[\mu\nu,\rho]} + \frac12
F_{[\mu\nu}A_{\rho]})\,. \ee The Einstein equation for $R_{66}$
gives: \be \nabla^2\phi = \frac{2\alpha}3e^{8\alpha\phi}
H^{\tau\rho\sigma}H_{\tau\rho\sigma} + \alpha e^{4\alpha\phi}
F^{\rho\sigma}F_{\rho\sigma} \,, \ee which is consistent with the
equations of motion following from the following five-dimensional
EMDA Lagrangian
 \be\lb{O435}
{\cal L}^5_{EMDA} = \sqrt{g\5}\left(R - \frac12(\partial\phi)^2 -
\frac14e^{4\alpha\phi}F_{\mu\nu}F^{\mu\nu}
-\frac1{12}e^{8\alpha\phi}H_{\mu\nu\lambda}H^{\mu\nu\lambda}\right)\,.
 \ee

\setcounter{equation}{0}
\section{Truncations in $D=5$ }
Consistent truncations in five dimensions are governed by the double
chain shown in (\ref{chain5}). The right subchain is rather
straightforward: to go from $EM_2DA$ to $EMDA$ and then to $EDA$
amounts to dropping the first and the second vectors, and further
dropping of the two-form leads to $VG$. The left subchain $ O(4,4)
\longrightarrow O(4,3) \longrightarrow G_{2(2)} \longrightarrow
SL(3,R)\,$ is not so trivial. In terms of  theories this means
 \be
U(1)^3 SG \longrightarrow EMDA \longrightarrow MSG\longrightarrow
VG\,.
 \ee
Thus $D=5\; U(1)^3$ supergravity, whose $D=3$ reduction is the coset
$O(4,4)/O(4)\times O(4)$, can be consistently truncated to minimal
$D=5$ supergravity via an intermediate $D=5$ $EMDA$ theory (an
oxidation point of the coset $O(4,3)/(O(4)\times O(3)$ ). The direct
truncation $U(1)^3 SG \longrightarrow MSG$ is in fact much simpler
and amounts to the identifications
 \be
X_1=X_2=X_3=1,\quad A^1=A^2=A^3.
 \ee
However, passing through the intermediate $EMDA$ step gives further
dualities in our list of theories. So, first we make the
identifications (following from (\ref{trunc44to43}))
 \be
X_1=X_2\equiv X,\quad X_3=\fr1{X^2},\quad A^1=A^2\equiv
\fr1{\sqrt{2}} A,
 \ee
arriving at the EMDA theory (\ref{O435}). The field equations
deriving from this Lagrangian are:
 \ba
&&D_{\rho}(e^{8\alpha\phi}H^{\mu\nu\rho}) = 0\,, \lb{H}\\
&&D_{\nu}(e^{4\alpha\phi}F^{\mu\nu} +
e^{8\alpha\phi}H^{\mu\nu\rho}A_{\rho}) =
0\,, \lb{F} \\
&&\nabla^2\phi = \alpha e^{4\alpha\phi}F_{\mu\nu}F^{\mu\nu}
+ \frac{2\alpha}3e^{8\alpha\phi}H_{\mu\nu\lambda}H^{\mu\nu\lambda}\,, \lb{phi}\\
&&R_{\mu\nu} - \frac12Rg_{\mu\nu} = \frac12\left\{\partial_{\mu}\phi
\partial_{\nu}\phi + e^{4\alpha\phi}F_{\mu\lambda}{F_{\nu}}^{\lambda}
+ \frac12e^{8\alpha\phi}H_{\mu\lambda\rho}{H_{\nu}}^{\lambda\rho} \right.\nn \\
&& \quad \left. - g_{\mu\nu}\left[\frac12\partial_{\lambda}\phi
\partial^{\lambda}\phi + \frac14e^{4\alpha\phi}F_{\lambda\rho}F^{\lambda\rho}
+
\frac1{12}e^{8\alpha\phi}H_{\lambda\rho\sigma}H^{\lambda\rho\sigma}
\right] \right\}\,, \lb{E}
 \ea
together with the consistency conditions following from the
definitions of the two- and three-forms,
 \ba
&&D_{\rho}\tilde{F}^{\mu\nu\rho} = 0\,, \lb{tF}\\
&&D_{\nu}(\tilde{H}^{\mu\nu} -
\frac12\tilde{F}^{\mu\nu\rho}A_{\rho}) = 0\,. \lb{tH}
 \ea
Equation (\ref{H}) is solved by
 \be H^{\mu\nu\rho} =
\frac1{\sqrt2}e^{-8\alpha\phi}\tilde{K}^{\mu\nu\rho}\,, \quad
(K_{[\sigma\tau,\lambda]} = 0)\,.
 \ee
Inserting this in (\ref{F}), (\ref{tH}) and (\ref{phi}) gives
 \ba
&&D_{\nu}\left(e^{4\alpha\phi}F^{\mu\nu} + \frac1{\sqrt2}
\tilde{K}^{\mu\nu\rho}A_{\rho}\right) = 0\,, \\
&&D_{\nu}\left(e^{-8\alpha\phi}K^{\mu\nu} + \frac1{\sqrt2}
\tilde{F}^{\mu\nu\rho}A_{\rho}\right) = 0\,, \\
&&\nabla^2\phi = \alpha\left(e^{4\alpha\phi}F_{\mu\nu}F^{\mu\nu} -
e^{-8\alpha\phi}K_{\mu\nu}K^{\mu\nu}\right)\,,
 \ea
which are consistent with the constraints
 \be\lb{O43G2}
\phi = 0\,, \quad K_{\mu\nu} = \pm F_{\mu\nu}\,.
 \ee
The remaining field equations
 \ba
&&D_{\nu}\left(F^{\mu\nu} \pm \frac1{\sqrt2}
\tilde{F}^{\mu\nu\rho}A_{\rho}\right) = 0\,, \\
&&R_{\mu\nu} - \frac12Rg_{\mu\nu} = \frac34\left(
F_{\mu\lambda}{F_{\nu}}^{\lambda} -  \frac14g_{\mu\nu}
F_{\lambda\rho}F^{\lambda\rho}\right)\,
 \ea
can be transformed by the rescaling
 \be
F_{\mu\nu} = \pm\sqrt{\frac23}f_{\mu\nu}
 \ee
into the equations of minimal five-dimensional supergravity,
deriving from the Lagrangian
 \be \lb{5min}
{\cal L}^5_{MSG}  = \sqrt{g} \bigg(R -
\frac14f^{\mu\nu}f_{\mu\nu}\bigg) -
\frac1{12\sqrt3}\epsilon^{\mu\nu\rho\sigma\lambda}f_{\mu\nu}
f_{\rho\sigma}a_{\lambda}\,.
 \ee
The three-dimensional invariance group is $G_{2(2)}$. This can be
further truncated to five-dimensional vacuum gravity, with
three-dimensional group $SL(3,R)$, by setting the 2-form $f$ to
zero.

\setcounter{equation}{0}
\section{Truncations in four dimensions}
The four-dimensional theories in the list (\ref{list}) can be
obtained by truncation of two compactified  five-dimensional
theories via several chains:
 \be\lb{chain4}
\begin{array}{cc} \hspace{-.7cm} \phantom{D=5} \;\;\qquad O(5,3)\quad  O(4.4)\quad
 \\ \;\;\;\;\;\; \swarrow \quad \searrow \quad \swarrow\qquad \\
\hspace{-1 cm} \phantom{D=4}\quad O(5,2)\quad O(4,3)\qquad \qquad \\
 \quad \searrow \quad \swarrow\;\;\qquad \searrow \qquad\\
  \quad SU(2,2) = O(4,2) \qquad O(3,3)= SL(4,R)\\
\hspace{-1.5 cm}  \downarrow\qquad\qquad\qquad \searrow \qquad\swarrow\qquad  \\
\hspace{-.8 cm}SU(2,1)\;\;\;\qquad\qquad O(3,2)\;\;\;\;\;\;\;  \qquad \qquad  \\
\hspace{-2.2cm}\downarrow \qquad\qquad \qquad\qquad\downarrow \\
\hspace{-.8  cm} SL(2,R)\;\;\;\quad\leftarrow\; \;\;\quad
O(2,2)\quad \quad\qquad\quad
  \end{array}\qquad \qquad\begin{array}{cc}    \qquad \{4A_\mu,\,4S,\,B_{\mu\nu}\}
   \quad \{4A_\mu,\,6S\}\quad
 \\ \;\;\; \swarrow \qquad \searrow \quad \swarrow\qquad \\
 \hspace{-1 cm} EM_3DA\qquad \;\;LLPT\qquad  \\
  \hspace{-1 cm}\searrow \quad \quad \swarrow\qquad \downarrow  \\
  \hspace{-.7 cm} EM_2DA \qquad CGMS\qquad\\
  \hspace{-.5 cm} \downarrow  \;\;\qquad\searrow \quad\downarrow   \\
  EM\qquad EMDA\;\;\;\;\;\;\;  \qquad   \\
 \hspace{-.6 cm} \downarrow \qquad  \qquad\downarrow \\
   \quad\;\;\; VG\;\;\;\leftarrow \;\;\; EDA\quad  \qquad\;\;\;
  \end{array}
 \ee
The first line indicates four-dimensional oxidation points of the
cosets of $O(5,3)$  and $O(4,4)$, these are beyond the list
(\ref{list}). Lower lines correspond to different truncation schemes
of these. In what follows we concentrate on the $O(4,3)$ subchain
and do not give details of the subchain going through $O(5,2)$.
There is also another chain leading to four-dimensional
Einstein-Maxwell theory via $D=5$ minimal supergravity, this will be
discussed below separately.

\subsection{$ O(4,3)$ theory}
Reduction of the $O(4,3)$ model from five to four dimensions leads
to the Lagrangian obtained by Lavrinenko, Lu, Pope and Tran (LLPT)
\cite{lav}
 \ba
{\cal L}_{LLPT} &=& \sqrt{g\4}\left(R -
\frac12\left[(\partial\phi)^2 + (\partial\varphi)^2 +
e^{2\phi}(\partial\chi)^2 +
e^{\sqrt2\varphi}(\partial\sigma)^2\right]\right. \nn\\
&&  -
\frac14e^{-\phi}\left[e^{\sqrt2\varphi}F^{+}_{\mu\nu}F^{+\mu\nu}
 + F^{0}_{\mu\nu}F^{0\mu\nu} + e^{-\sqrt2\varphi}F^{-}_{\mu\nu}F^{-\mu\nu}
\right] \nn\\
&& \left.- \frac12\chi\left[\tilde{F}^{+}_{\mu\nu}F^{-\mu\nu} +
\frac12\tilde{F}^{0}_{\mu\nu}F^{0\mu\nu}\right]\right)\,,
 \ea
with three Maxwell fields
 \ba
F^{-}_{\mu\nu} &=& 2A^-_{[\nu,\mu]} \,, \nn\\
F^{0}_{\mu\nu} &=& 2\left(A^0_{[\nu,\mu]} + \sigma
A^-_{[\nu,\mu]}\right)\,,
\nn\\
F^{+}_{\mu\nu} &=& 2\left(A^+_{[\nu,\mu]} - \sigma A^0_{[\nu,\mu]} -
\frac12\sigma^2 A^-_{[\nu,\mu]}\right)\,.
 \ea
This theory can be truncated to $D=4$ $EM_2DA$ with two vector
fields, which has the hidden symmetry  $O(4,2)\sim SU(2,2)$.

The field equations for the dilaton $\varphi$ and the axion $\sigma$
are
 \ba\lb{LLPT}
\nabla^2\varphi &=& \frac1{\sqrt2}\varphi \left[e^{\sqrt2\varphi}
(\partial\sigma)^2 \right.\nn\\
&& \left.\quad + \frac12e^{-\phi}\left(
e^{\sqrt2\varphi}F^{+}_{\mu\nu}F^{+\mu\nu}
- e^{-\sqrt2\varphi}F^{-}_{\mu\nu}F^{-\mu\nu}\right)\right] \,, \\
D_{\mu}\left[e^{\sqrt2\varphi}\partial^{\mu}\sigma\right] &=&
F^0_{\mu\nu}\left[e^{\sqrt2\varphi-\phi}F^{+\mu\nu} -
e^{-\phi}F^{-\mu\nu}\right]\,.
 \ea
These equations are consistent with the identifications
 \be
\varphi = 0\,, \quad \sigma = 0\,, \quad F^{-}_{\mu\nu} =
F^{+}_{\mu\nu}\,,
 \ee
while the equations for the two-forms, written for $\varphi = \sigma
= 0$,
 \ba
D_{\nu}\left[e^{-\phi}F^{+\mu\nu} + \chi\tilde{F}^{-\mu\nu}\right] &=& 0\,, \\
D_{\nu}\left[e^{-\phi}F^{0\mu\nu} + \chi\tilde{F}^{0\mu\nu}\right] &=& 0\,, \\
D_{\nu}\left[e^{-\phi}F^{-\mu\nu} + \chi\tilde{F}^{+\mu\nu}\right]
&=& 0
 \ea
are also consistent with the identification of the two-forms $F^+$
and $F^-$. The truncated Lagrangian \ba\lb{O42} {\cal L}\4 &=&
\sqrt{g\4}\left(R - \frac12\left[(\partial\phi)^2 +
e^{2\phi}(\partial\chi)^2 + \right]\right. \nn\\
&&  \left. - \frac14\left[e^{-\phi}F^{a}_{\mu\nu}F^{a\mu\nu} +
\chi\tilde{F}^{a}_{\mu\nu}F^{a\mu\nu}\right]\right) \ea (with $F^1
\equiv F^0$ and $F^2 \equiv \sqrt2 F^{\pm}$ is that of the  $EM_2DA$
model \cite{Gal'tsov:1996cm,Gal'tsov:1997kp}, which after reduction
to three dimensions has the symmetry $O(4,2) \sim SU(2,2)$. $EM_2DA$
can then be truncated  to $p=1$ $EMDA$ (hidden symmetry group
$O(3,2) \sim Sp(4,R)$) by identifying the two 2-forms (or setting
one of them to zero) \cite{Gal'tsov:1997kp}. In turn, $EMDA$ can be
truncated to four-dimensional $EDA$, with three-dimensional group
$O(2,2) \sim SL(2,R)\times SL(2,R)$, by setting the remaining
two-form to zero, and finally to four-dimensional vacuum gravity.

$EM_2DA$ can also be truncated   to four-dimensional
Einstein-Maxwell theory, with three-dimensional invariance group
$SU(2,1)$ by the consistent constraints
 \be \phi = 0\,, \quad \chi =
0\,, \quad F^{2}_{\mu\nu} = \tilde{F}^{1}_{\mu\nu} \,.
 \ee
 Finally, Einstein-Maxwell theory can be truncated to
four-dimensional vacuum gravity (three-dimensional group $O(2,1)
\sim SL(2,R)$).

\subsection{CGMS subchain }

Another chain of reduction/truncations shown in (\ref{chain4}) goes
from $LLPT$ to the theory obtained by Chen, Gal'tsov, Maeda and
Sharakin \cite{Chen:1999rv} (CGMS) as a compactification of the
five-dimensional $EDA$, which in turn is a compactification of
six-dimensional vacuum gravity \cite{Clement:2001gia}). This amounts
to setting
 \be
\varphi = \psi/\sqrt{2}\,, \quad \sigma = 0\,, \quad  F^{+}_{\mu\nu}
={\cal F}_{\mu\nu}\,,\quad F^{-}_{\mu\nu}=F_{\mu\nu}\,, \quad
F^0_{\mu\nu}=0\,, \ee which is also consistent with the equations of
motion (\ref{LLPT}). The resulting lagrangian is \cite{Chen:1999rv}:
 \be
{\cal L}_{CGMS}  =  \sqrt{g\4}\left(R - \frac12
(\partial\phi)^2-\frac12 e^{2\phi}(\partial\chi)^2  -
\frac14(\partial\psi)^2
  - \frac14e^{-\phi}\left[e^{ \psi}{\cal F}_{\mu\nu}{\cal
F}^{\mu\nu}
  + e^{-\psi}F_{\mu\nu}F^{\mu\nu}
\right] - \frac{\chi}2 \tilde{F}_{\mu\nu}{\cal F}^{\mu\nu}
 \right)\,.
 \ee
Subsequent truncation to $EMDA$ with one vector field amounts to
setting
 \be
\psi=0\,,\quad  {\cal F}_{\mu\nu}=F_{\mu\nu}\,.
 \ee
 The remaining segment $EMDA\longrightarrow EDA \longrightarrow VG$
is the same as in the previous subsection. This subchain does not
include Einstein-Maxwell theory.

\subsection{From $G2$ to EM}

Meanwhile there is another reduction/truncation chain from five to
four dimensions, not shown on the table (\ref{chain4}), leading to
$D=4$ Einstein-Maxwell theory. Namely, it can also be obtained by
truncation of minimal five-dimensional supergravity reduced to four
dimensions since $SU(2,1) \subset G_{2(2)}$: the corresponding
sequence is
 \be\label{g2chain}
O(4,3) \longrightarrow G_{2(2)} \longrightarrow SU(2,1)
\longrightarrow O(2,1) \,.
 \ee
One starts with the five-dimensional action (\ref{5min}). Assuming
the existence of a space-like Killing vector $\partial_z$, one can
parametrize the five-dimensional metric and the Maxwell field $A\5$
by
\begin{eqnarray}
ds\5^2 &=& \mathrm{e}^{-2\phi } (dz + C_\mu dx^\mu)^2 +
\mathrm{e}^{\phi} ds\4^2,\label{ds5}\\
A\5 &=& A_\mu dx^\mu + \sqrt{3}\kappa dz\,.
\end{eqnarray}
The corresponding four-dimensional lagrangian reads
\begin{equation}
{\cal L}_4 =   \sqrt{ g_4} \left[ R - \frac32 (\partial \phi)^2 -
\frac32 \mathrm{e}^{2\phi} (\partial \kappa)^2 - \frac14
\mathrm{e}^{-3\phi} G^2 - \frac14\mathrm{e}^{-\phi} \bar F^2 -
\frac12 \kappa F {\tilde F} \right],
\end{equation}
where
\begin{equation}
 G = dC, \quad F = dA, \quad \bar F = F + \sqrt{3} C \wedge d\kappa,
\end{equation} and $\tilde F$ is the four-dimensional Hodge
dual of $F$. This describes an Einstein theory with two coupled
abelian gauge fields $F$ and $G$, a dilaton $\phi$ and an axion
$\kappa$. The field equations in terms of the four-dimensional
variables read
\begin{eqnarray}
\nabla^2 \phi - \mathrm{e}^{2\phi} (\partial \kappa)^2 + \frac14
\mathrm{e}^{-3\phi} G^2 + \frac1{12} \mathrm{e}^{-\phi} \bar F^2 &=&
0,
\\
\nabla_\mu \left( \mathrm{e}^{2\phi} \nabla^\mu \kappa \right) -
\frac1{3} \left[ \sqrt3\,  \nabla_\mu (\mathrm{e}^{-\phi} \bar
F^{\mu\nu} C_\nu) + \frac12 F_{\mu\nu} {\tilde F}^{\mu\nu} \right]
&=& 0,
\\
\nabla_\mu \left( \mathrm{e}^{-\phi} \bar F^{\mu\nu} + 2\kappa
{\tilde F}^{\mu\nu} \right) &=& 0,
\\
\nabla_\mu \left( \mathrm{e}^{-3\phi} G^{\mu\nu} \right) +  \sqrt3\,
\mathrm{e}^{-\phi} \bar F^{\mu\nu} \partial_\mu \kappa &=& 0\,.
\end{eqnarray}
The consistent truncation to the Einstein-Maxwell theory  is
achieved via the constraints
 \be
\phi = 0\,, \quad \kappa = 0\,, \quad G_{\mu\nu} = \frac1{2\sqrt3}
\sqrt{ g_4}\epsilon_{\mu\nu\rho\sigma}F_{(4)}^{\rho\sigma}\,.
 \ee
Then the first two equations are trivially satisfied, while the last
two become the Maxwell equation and Bianchi identity for
$F_{\mu\nu}$, the Lagrangian reducing to that of the $EM$ theory.

In a recent paper \cite{seven} we have derived a new simple matrix
representative of the coset $O(4,3)/O(4)\times O(3)$, which can also
be used in the case of minimal five-dimensional supergravity. The
truncations (\ref{g2chain}) are implemented there, imposing
algebraic constraints on the potentials.

It is worth noting that the $D=3$ $EM$ coset $SU(2,1)/S(U(2)\times
U(1))$ has $D=4$ as an oxidation endpoint \cite{Cremmer:1999du}, so
that this theory cannot be obtained by toroidal dimensional
reduction from higher-dimensional theories. However, as we have
shown, it can be embedded into higher-dimensional theories using
reduction/truncation chains.

\section{Conclusions}
We have described various mappings between six, five and
four-dimensional theories whose $D=3$ cosets are subspaces of
$O(5,4)/(O(5)\times O(4))$. Among them are truncated heterotic
effective actions ($EMDA$), minimal supergravities in five and six
dimensions, $U(1)^3$ five-dimensional supergravity and some popular
four-dimensional models. These mappings open  a way to construct
exact solutions in five and (especially) in six dimensions starting
from known four-dimensional solutions. For this purpose we have
given a detailed description of reduction/truncation chains in term
of natural non-reduced field variables. Using them one can easily
uplift lower-dimensional solutions to higher dimensions. Combining
this with coset transformations and generation of solutions as
geodesics of coset spaces, one acquires new tools for solution
generation. Applications will be given elsewhere.

\section*{Acknowledgments} The authors are grateful to
Chiang-Mei Chen and Nicolai Scherbluck for fruitful collaboration.
D.G. is grateful to the LAPTH Annecy for hospitality in December
2012 while the paper was written. His work was supported in part by
the RFBR grant 11-02-01371-a.

\end{document}